\documentclass{article}
\usepackage{graphicx}\usepackage{amsfonts}
\begin{document}
\title{
\vspace{2cm} \bf{Chiral symmetry and strangeness content in
nuclear physics parametrized by a medium dependent bag constant }}
\author{R. Aguirre\\{\it Depto. de Fisica, Facultad de Ciencias Exactas,}\\
{\it Univ. Nacional de La Plata, Argentina}}
%\address{Depto. de Fisica, Univ. Nacional de La Plata}
%\date{ }
 \maketitle
\begin{abstract}
Non-perturbative QCD vacuum effects at finite density are
parametrized by means of a bag constant $B$. It is extracted from
a Nambu- Jona Lasinio model with two or three flavors. The
parameter $B$ is used in an effective bag-like model of baryons to
study the nuclear phenomenology. We examine the nucleon structure
and the thermodynamical properties of symmetric nuclear matter,
particular attention is paid to the symmetry energy and to the
eventual phase transition to deconfined quark matter. An
alternative sketch of the binding mechanism of symmetric nuclear
matter emerges within this approach. It is also found that the
inclusion of strangeness content in $B$ is crucial for an
appropriate description.
\end{abstract}
\vspace{1cm}

 The low energy regime of QCD is very difficult to
manage due to its non-perturbative nature. This feature gives rise
to different methods, like sum rules and effective models, to
avoid dealing with the full theory in the description of the
hadronic phenomenology. These approaches intend to include the
symmetries of QCD implicitly into the final results.

A simple shortcut procedure to take care of non-perturbative
vacuum contributions is the introduction of the so-called bag
constant $B$, in both confined and deconfined quark phases. The
former bag models used it to account for the energy required to
create a unit volume of the Weyl-Wigner vacuum into the complex
QCD ground state. In ref.\cite{ADAMI} two different scales were
associated to $B$, the higher one is related to the condensation
of gluons and consequently with the recovery of the scale symmetry
of QCD. The lower one is indicative of the restoration of chiral
symmetry. Approximate calculations for these scales there given
are $B^{1/4}=245 MeV$ and $B^{1/4}=140 MeV$ respectively.

Another significant aspect in the definition of the bag parameter
is its hypothetical variation with the medium properties of the
hadronic matter. Speculation about the medium dependence of $B$
has been motivated from different areas and ad-hoc
phenomenological parametrizations have been proposed \cite{JIN,
A&S, LU, BLASCHKE, BURGIO}. It is expected that the gradual
recovery of fundamental symmetries with increasing density and/or
temperature modifies the vacuum structure and the value of $B$
accordingly.

In this work  we attempt to extract a bag parameter that varies
with the baryonic density but carrying the signature of the chiral
symmetry and, ultimately to examine its influence over	the
nuclear dynamics at finite density. For this purpose we use two
rather different theoretical descriptions of the high density
regime of the strong interacting matter and we combine them to
describe in a simplified manner the influence of the QCD vacuum
changes over the gross properties of the nuclear matter. We have
selected the Nambu-Jona Lasinio (NJL) model \cite{NJL0} to study
the modification of the QCD vacuum with increasing baryonic
density and to extract a medium-dependent bag constant. This
parameter is then inserted into the Quark Meson Coupling (QMC)
model \cite{QMC0,QMC1} to describe the nucleon substructure. It
must be mentioned that each one of these issues have been
independently investigated in the past, but we intend to put them
together in a novel combination. For instance, a density dependent
bag constant has been proposed within the QMC \cite{JIN}, but it
was initially regarded as a way to restore certain aspects of the
nuclear phenomenology instead of a manifestation of the underlying
QCD vacuum. This fact explains why the parametrization of $B$ has
been made in terms of either the scalar $\sigma$ meson	mean field
or the effective nucleon mass.

Adopting the viewpoint of the early bag models, we expect that $B$
would carry significative information about the fundamental
symmetries and would transfer it to the hadronic sector. Thus, as
previously mentioned, the recovery of chiral and scale symmetries
with increasing density and/or temperature would have remarkable
effects over $B$.

The formulation of chiral invariant theories was certainly one of
the major successes in the development of the bag models
\cite{THOMAS}. However QMC abandoned this requirement since it was
thought as a way to match the bag picture of the hadron structure
with the Quantum Hadrodynamics \cite{QHD} treatment of the nuclear
many-body problem. The coupling to a scalar $\sigma$ meson plays a
key role in this description, therefore the explicit breaking of
chiral symmetry seems to be an inherent feature of the QMC
formulation. In the following we shall see how a bag constant
respecting the chiral invariance in its definition, could modify
this situation.

Our starting point is the  original  meaning of  $B$, i.e. the
energy difference between the fundamental state with explicit
realization of the symmetries and that with symmetries
spontaneously broken \cite{BHADURI}. For this purpose we consider
the NJL for both SU(2) and SU(3) versions of the chiral symmetries
\cite{NJLSU3, REHBERG, KLEVANSKY}. In the last case the flavor
degeneracy is removed introducing a heavy strange quark.

The lagrangian density of the model is \cite{KLEVANSKY}
\begin{equation}
{\cal L}=\bar{q}(i\slash \! \! \! \partial-m)q+{\cal L}_4+{\cal
L}_6
\end{equation}
with $q^t=(u,d,s)$ and	$m=$diag$(m_u,m_d,m_s)$ is the current
mass matrix that explicitly breaks chiral invariance. The four
quarks interaction term ${\cal L}_4$ is invariant under $U_L(3)
\times U_R(3)$ transformations
\begin{equation}
{\cal L}_4=G \left[ (\bar{q} \lambda_k q )^2 + (\bar{q}i \gamma_5
\lambda_k q )^2\right],
\end{equation}
where $\lambda_k$, $k=0,...,8$ are the Gell-Mann matrices with
$\lambda_0=\sqrt{2/3}$ diag$(1,1,1)$.

The six quarks term removes the redundant $U_A(1)$ symmetry
\begin{equation}
{\cal L}_6=-K  \left\{ \det[\,\bar{q}_i(1+\gamma_5)q_j]+
\det[\,\bar{q}_i (1-\gamma_5)q_j] \right\}
\end{equation}

 Due to the development of the quark-antiquark condensates the
current quark masses acquires its large constituent values $M_k$,
given by the self-consistent equations \cite{KLEVANSKY}
\begin{equation}
M_k=m_k-4G <\bar{q}_k q_k>+2K<\bar{q}_i q_i><\bar{q}_j q_j>,
\label{GAP}
\end{equation}
with $i\neq k \neq j$, and the condensates are given by
\begin{eqnarray}
<\bar{q}_k q_k>&=&C_k=-i \lim_{x'_0 \rightarrow x_0+}\int \frac{d^4p}{(2\pi)^4}
\,e^{-i p.(x'-x)}\, {\mathnormal Tr} G_k(p)\\
&=&-\frac{N_c}{2\pi^2}M_k\left[ \Lambda E_\Lambda- p_k E_{p_k}	-
M_k^2 \ln(\frac{\Lambda+E_\Lambda}{p_k+E_{p_k}})\right],
\label{CONDENSATE}
\end{eqnarray}
where we have used the quark propagator
\begin{equation}
G_k(p)=(\slash \!\!\!\! p+M_k)\left[\frac{1}{p^2-M_k^2+i
\epsilon}+2 \pi i \,\delta(p^2-M_k^2)\,\theta(p_k-p) \right]
\end{equation}
and $E_x=\sqrt{x^2+M_k^2}$. We have used a non-covariant
3-momentum cutoff $\Lambda$, and explicitly included finite
density effects using the Fermi momentum $p_k$. Therefore the
ground state is obtained  filling the quark Fermi shell plus the
Dirac sea divergent contribution. The contribution of $k$-flavor
$n_k$ to the total baryon density is related to its Fermi momentum
by

\begin{equation}
n_k=\frac{N_c}{3}\frac{p_k^3}{3 \pi^2}.
\end{equation}

Once the eqs. (\ref{GAP}) and (\ref{CONDENSATE}) have been solved
the energy density for the fundamental state is evaluated by
taking the statistical average of the hamiltonian density
\begin{equation}
\varepsilon_{QM}=\frac{N_c}{8 \pi^2}\sum_k (F(M_k,p_k)-
F(M_k,\Lambda))+2G(C_u^2+C_d^2+C_s^2)-4K  C_u C_d C_s,
\label{NJLENERGY}
\end{equation}
with $F(M,x)=x E_x (E_x^2+x^2)-M^4 \ln((x+E_x)/M)$.

A definite constant $\varepsilon_0$ must be added to Eq.
(\ref{NJLENERGY}) in order to get zero energy when no valence
quarks are present. The pressure is evaluated using the
thermodynamical relation
\begin{equation}
P_{QM}=\sum_k \mu_k n_k-\varepsilon_{QM},
\end{equation}
where $\mu_k=\partial \varepsilon_{QM}/\partial n_k$ is the
chemical potential for quarks of  flavor $k$.

As above mentioned we take $B$ as the difference between Eq.
(\ref{NJLENERGY}) and the energy of the state with zero quark
condensates, but at the same baryonic density. This definition is
in agreement with that used in \cite{BUBALLA}. In this way a
density dependent bag parameter has been obtained, able to be used
in both hadron and quark matter since the parametrization is made
in terms of only a common conserved charge. It is expected that
the flavor composition of matter affects the density dependence of
$B$, therefore we examine three different situations (a) symmetric
$u-d$ quark matter in the SU(2) version of the model, (b)
symmetric $u-d$ quark matter in the SU(3) version, that is the
Dirac sea contribution of strange quarks is included, and (c)
electrically neutral quark matter coexisting with leptons in
equilibrium by electroweak decay, a situation of interest in
astrophysical applications. Thus we have the constraints for the
quark Fermi momenta $p_u=p_d=p, p_s=0$, $n=2N_cp^3/(9\pi^2)$ for
the items (a)-(b) and $\mu_s=\mu_d, \mu_u+\mu_e=\mu_d$,
$0=N_c(2p_u^3-p_d^3-p_s^3)/3-p_e^3$, $n=N_c (p_u^3+p_d^3+p_s^3)/(9
\pi^2)$ for the case (c). Here $n$ stands for the baryon number
density and $\mu_e$ for the electron chemical potential.

For practical calculations we have used $m_u=m_d=5.5 MeV,
m_s=140.7 MeV$ and the model parameters $\Lambda=602.3$ MeV, $G
\Lambda^2=1.835$, $K \Lambda^5=12.36 MeV$ which were obtained by
reproducing the $\pi$-, $K$- and $\eta$'-meson masses together
with the constant $f_\pi$ \cite{REHBERG}, for cases (b) and (c).
For the item (a) instead, we have used the parameters given in
\cite{KLEVANSKY} for the Hartree results in the 3-momentum cutoff
approach, i.e. $\Lambda=653 MeV$, $G \Lambda^2=2.14$.

 The density dependence obtained in our calculations is shown in
Fig.1. The SU(2) and SU(3) schemes differ essentially in all the
range, while within the three flavor treatment the cases (b) and
(c) are similar for low and medium densities. The items (a) and
(c) exhibit a steep descent at sufficiently high densities, in
contrast with case (b) that keeps a large constant value. The
cause of this behavior is that the strange condensate does not
melt in (b) as the non-strange ones do. In turn this is due to the
lack of the valence strange quark contribution, as confirmed by
the sudden splitting of the curves (b) and (c) at $n\simeq 4.5\,
n_0$ shown in Fig. 1. The fast decrease of the graph corresponding
to (c) coincides with the turning up of the strange quark into the
Fermi sea. Therefore the large asymptotic value shown by the SU(3)
approach for the symmetric $u-d$ matter can be assigned to the
constraint of null strange Fermi momentum.

The resulting $B$ parameter can be used to evaluate nuclear matter
properties. For this purpose we have selected a bag-like model
that provides a satisfactory description of a wide range of
nuclear phenomena, the QMC model \cite{QMC0,QMC1}. Within the QMC
baryons are regarded as non-overlapping spherical bags, where
three valence quarks are confined. These quarks interact with its
surrounding media by the interchange of light $\sigma$ and
$\omega$-mesons, as suggested by the successful Quantum
Hadrodynamics models \cite{QHD}  . Inside the bag the quark fields
obey the mean-field equation
\begin{equation}
( i \gamma^{\mu} \partial_{\mu} - g_{\omega}^k \gamma^0 \omega_0 -
{m_k}^\ast) q_k(x) = 0, \label{QMCEQ}
\end{equation}
where $m_k^\ast=m_k-g_\sigma^k \sigma$, since within the bag there
is only a  small breakdown of chiral symmetry we use the current
value of the quark masses $m_k$. In Eq. (\ref{QMCEQ}) $\sigma$ and
$\omega_0$ stand for the mean field values of the meson fields
evaluated in the dense hadronic medium and $g_{\sigma, \omega}^k$
are the quark-meson coupling constants. The solutions of this
equation is \cite{QMC1}
\begin{equation}
q_k(r,t)={\cal N}^{-1/2} \frac{e^{-i \varepsilon_k t}}
{\sqrt{4\pi}} \left( \begin{array}{c}
j_0 (x_k \, r/R) \\
i \beta_k {\vec{\sigma}}.{\hat{r}} j_1 (x_k \, r/R)
\end{array} \right) \chi ^k,
\end{equation}
where $\chi ^k $ is the quark spinor and
\begin{eqnarray}
\varepsilon_k &=& \frac{\Omega_k}{\!R} + g_\omega^k \;\omega_0,
\nonumber \\
{\cal N }&=&R^3\;[2 \Omega_k (\Omega_k - 1) + R {m_k}^\ast ]\,
\frac{ j_0^2 (x_k) }{x_k^2} \nonumber \\
\beta_k&=&{\left[ \frac{\Omega_k - R m_k^\ast }{\Omega_k + R
m_k^\ast } \right]}^{1/2}
\end{eqnarray}
with $\Omega_k =[x_k^2 +(R m_k^\ast)^2	  ]^{1/2}$. The eigenvalue
$x_k$ is solution of the equation
\begin{equation}
j_0 (x_k) = \beta_k \; j_1 (x_k) \label{BOUNDARY}
\end{equation}
which arises from the condition of zero quark current through the bag surface.\\
In this model the ground state bag energy is identified with the
baryon mass $M^\ast$,
\begin{equation}
M^\ast=\frac{\sum_k N_k \Omega_k - z_0}{R} + \frac{4}{3} \pi B R^3
\label{BAGMASS}
\end{equation}
where $N_k$ is the number of quarks of flavor $k$ inside the bag.
As usual in the bag model, two phenomenological corrective terms
are added to the quark energy: the one containing $z_0$ amends the
single particle spectrum and the other, proportional to the bag
volume, represents the energy required to create a Weyl-Wigner
bubble in the complex QCD vacuum. The last one has acquired a
density dependence in our approach.

The bag radius adjusts to maintain the confinement volume in
equilibrium with it surrounding media. Thus once $B$ is given and
$z_0$ fixed to reproduce the baryon mass at zero baryon density,
the bag radius $R$ is determined by the equilibrium condition
$dM^\ast/dR=0$

Density dependent bag parameters have been proposed previously
within the QMC description \cite{JIN,A&S,LU}. Formerly it was
regarded as a way to restore the phenomenology of $\sigma-\omega$
fields at the saturation nuclear density $n_0=0.15 fm^{-3}$
\cite{JIN}. However the density dependence of $B$ was guessed by
means of the effective nucleon mass or by the $\sigma$-mean field,
thus its relation with the underlying theory was not clear.
Instead a phenomenological relation between $B$ and the equation
of state of nuclear matter was the proposal in \cite{A&S}, and
restriction to experimental electron scattering analysis was
imposed in \cite{LU}. The decrease of $B$ at $0.7 n_0$ is of $55
\%$ and $22 \%$ for curves (a) and (b)-(c) respectively. In the
last case it is only a few percent above the upper bound of $10-17
\%$ established in \cite{LU}. Based on different grounds, the
necessity of a variable bag parameter was claimed in \cite{BURGIO}
in order to describe appropriately the structure of neutron stars.

Next we shall examine the behavior of symmetric nuclear matter at
medium and high densities under the varying $B$-parameter that
includes the signatures of the chiral symmetry of the strong
interaction, considering two or three flavors contributions.

The thermodynamical properties in the QMC model are evaluated
regarding nucleons as quasi-particles with single particle
spectrum $\epsilon_p=\sqrt{p^2+M^{\ast \, 2}}+g_\omega \omega_0$,
where the effective nucleon mass must be evaluated from Eq.
(\ref{BAGMASS}). The energy density and the pressure for symmetric
nuclear matter are, therefore \cite{QMC1}
\begin{eqnarray}
\varepsilon_{NM}&=&\frac{1}{4\pi^2}F(M^\ast,p_F)
% \left[ p_F
%\sqrt{p_F^2+M^{\ast \, 2}}(2p_F^2+M^{\ast \, 2})-4 M^{\ast \, 4}
%\sinh^{-1}(p_F/M^\ast)
% \right]
%\nonumber \\
% &&
 +\frac{1}{2}(m_\sigma^2 \sigma^2+m\omega^2 \omega^2), \label{NUCENER}\\
 P_{NM}&=&n \epsilon_{p_F} - \varepsilon_{NM}.
\end{eqnarray}
In Eq. (\ref{NUCENER}) the energy of mesons has been included,
$p_F$ is the nucleon Fermi momentum related with the baryon
density $n=2 p_F^3/(3 \pi^2)$, and conditions of homogeneity and
static equilibrium has been assumed. The symmetry energy of
nuclear matter is a very significative quantity to understand a
variety of nuclear processes at medium and high densities
\cite{BALI1}. For example, it has a crucial role in the dynamics
and structure of neutron stars, as well as in the composition of
matter in heavy-ion collisions. However its behavior far from the
normal nuclear density is poorly known experimentally, and
theoretical predictions range over a wide and sometimes
contradictory qualitative results. For a linear nucleon-$\rho$
meson coupling the energy symmetry for symmetric nuclear matter is
\begin{equation}
\epsilon_{sym}=\frac{1}{8}\left( \frac{g_\rho}{m_\rho} \right)^2
n+\frac{p_F^2}{6 \sqrt{p_F^2+M^{\ast \, 2}}}
\end{equation}

The mesons mean field values have not been determined yet, they
can be obtained  minimizing the energy density with respect to
$\sigma$ and $\omega$. This leads to the equations \cite{QMC1}
\begin{eqnarray}
\sigma&=&-\frac{1}{2 \pi^2 m_\sigma^2}\frac{\partial
M^\ast}{\partial \sigma} \left[ p_F \sqrt{p_F^2+M^{\ast \, 2}}-
M^{\ast \, 2} \sinh^{-1}(p_F/M^\ast)
 \right],
\\
\omega&=&\frac{g_\omega}{m_\omega^2}n.
\end{eqnarray}
The nucleon-meson couplings are expressed in terms of the
quark-meson couplings as $g_\sigma=3 g_\sigma^k, g_\omega=3
g_\omega^k$. Numerical values can be assigned  reproducing the
conditions of nuclear matter at the saturation density $n_0$.
Meaningfully it is found in the SU(3) treatment that, for a
certain value of the binding energy, one can do calculations
without the $\sigma$-meson coupling, i.e. $g_\sigma^k=0$. Thus we
choose the values $P_{NM}(n_0)=0$,
$\varepsilon_{NM}(n_0)-M_0=-16.807 MeV$ for the pressure and the
binding energy at the normal density. If the vector-isovector
meson $\rho$ is also considered, then the corresponding coupling
constant $g_\rho=g_\rho^k$ can be deduced  equating the symmetry
energy $\epsilon_{sym}=(\partial \epsilon_N/\partial n)$ at $n_0$
to its phenomenological value $\epsilon_{sym}=34 MeV$. The
coupling constants obtained under these assumptions are shown in
Table I.

Numerical calculations have been developed  taking $m_\sigma=2.787
fm^{-1}, m_\omega=3.968 fm^{-1}$, and $m_\rho=3.902 fm^{-1}$ for
the meson masses, $M_p=4.755 fm^{-1}, M_n=4.761 fm^{-1}$ for the
nucleon masses at zero baryon density. For the sake of comparison
we have also carried out calculations with the standard fixed $B$
treatment (in the following labelled case (d)).

Some results for the in-medium nucleon properties are displayed in
Fig.2. The neutron mass is uniformly decreasing, the falling at
$n_0$ is approximately $10\%$ in the SU(3) treatment, $20-25\%$ in
the constant $B$ option and of an excessively high $40 \%$ in the
SU(2) case. The behavior of the bag radius in terms of the baryon
density is shown in Fig. 2, the cases (b) and (d) provide a radius
with a bounded variation. In particular for the approach (b) an
increasing radius is obtained whose growth saturates $15 \%$ over
its zero density value. Instead for case (a) a monotonously
increasing radius is obtained, exceeding the limit of validity of
the model. The large increment of the bag radius yields a
violation of the non-overlapping bag hypothesis at $n\simeq n_0$,
thus we may conclude that the two flavor description of the bag
parameter is not able of true physical consideration.
 In the three flavor case
this assumption is also broken down but at the higher density $n
\simeq 6 \, n_0$.

 The results for the pressure and the symmetry
energy are shown in Fig. 3, it can be seen that the softest
equation of state at high densities  corresponds to (d). The
nuclear compressibility $\kappa=9 \,\partial P_{NM}/\partial n$
takes on the values $310$ and $335 MeV/fm^3$ at $n=n_0$ for the
(d) and (b) approaches, respectively. For the symmetry energy  all
the results are very similar for low densities and differences
becomes sensible only beyond $n/n_0\sim 2.5$.

Thermodynamical quantities are similar in the approaches (b) and
(d), corresponding greater pressures and lower symmetry energy to
the instance (b). Instead nucleon mass and radius have
distinguishing behaviors in all the range of densities.

The medium dependent $B$ description has an upper bound of
applicability given by the point where bags start to overlap,
unless new physical aspects arise before this situation is
reached. This could be the case if a deconfinement phase
transition take place. To examine this possibility we have
compared the equation of state obtained previously with that
corresponding to quark matter.	The Gibbs criterium for phase
transitions $P_{NM}=P_{QM}$ and $\mu_n=2 \mu_d+\mu_u$ is adopted,
and variable or fixed $B$ is used. Thus a first order phase
transition is found only for the case (a), but at so high density
that QMC model assumptions have been violated. Not transition at
all is found for the cases (b).

In this paper we have studied the effects of including the chiral
symmetry of QCD and the strange degree of freedom in the
definition of the bag parameter $B$. This has been obtained from a
Nambu- Jona Lasinio model with two or three flavors, at finite
baryon density. Since baryon number is a conserved charge of both
quark and hadron matter, the quantity $B$ obtained can be regarded
as  medium-dependent and applied to examine nucleon properties and
the nuclear matter equation of state. Thus our results for $B$ are
based on a dynamical effective model of QCD, in contrast with
previous calculations that used {\it ad hoc} parametrizations in
terms of the $\sigma$-meson mean field. The description in terms
of only non-strange quarks must be discarded, i.e. the bag
parameter must include the strange vacuum contribution even for
the description of ordinary nuclear matter. Within the SU(3)
scheme we have found that the behavior of $B$ strongly depends on
the composition of matter at medium and high densities.

For the hadronic phase we used the QMC model of confined quarks
coupled to mesons, that allows us to extract the confinement
volume medium-dependence. From the comparison of the items (b) and
(d) we conclude that the case (b) does not need of the $\sigma$
meson to appropriately describe the properties of nuclear matter.
Therefore the saturation mechanism differs from those of Quantum
Hadrodynamics, the attractive contribution is provided by the
interplay between many-body effects and the modification of the
hadronic phase vacuum. Avoiding a direct coupling between quarks
and the $\sigma$-meson opens new possibilities such as a chiral
extension of the QMC, or could eliminate the unpleasant features
reported in \cite{JENNINGS}.

We have not found any first order phase transition from symmetric
nuclear matter to quark matter, but this could be a failure of the
approach that could be mended including quark correlations between
neighboring bags in the high density regime of the hadronic phase
\cite{A&D}.

 \vspace{0.5cm} {\centerline{\bf Acknowledgements}}

\centerline {\small This work was partially supported by the
CONICET, Argentina.}

\newpage
\begin{table}
\centering
\begin{tabular}{r|ccccc}
Case & $g_\sigma$ & $g_\omega$ & $g_\rho$ & $B_0^{1/4}[MeV]$& $R_0 [fm]$ \\
\hline
(a)&11.97 & 16.46 & 8.43 & 165.06 & 0.83  \\
(b)& 0.00 &  6.00 & 9.53 & 217.59 & 0.57  \\
(d1)& 17.39 &  8.26 & 9.17 & 165.06 & 0.83  \\
(d2)&18.30 &  9.23 & 9.04 & 217.59 & 0.57  \\
\end{tabular}
\caption{Coupling constants for the QMC model for the cases
considered: (a) and (b) correspond to medium dependent $B$
obtained from a Nambu-Jona Lasinio model with SU(2) and SU(3)
chiral symmetries respectively,  (d1) and (d2) correspond to the
constant $B$ fixed at the zero density value of (a) and (b)
respectively. In the last two columns the zero density value of
the bag parameter and the bag radius are shown. }
\end{table}

\newpage

\begin{figure}
\centering
\includegraphics[scale=0.7]{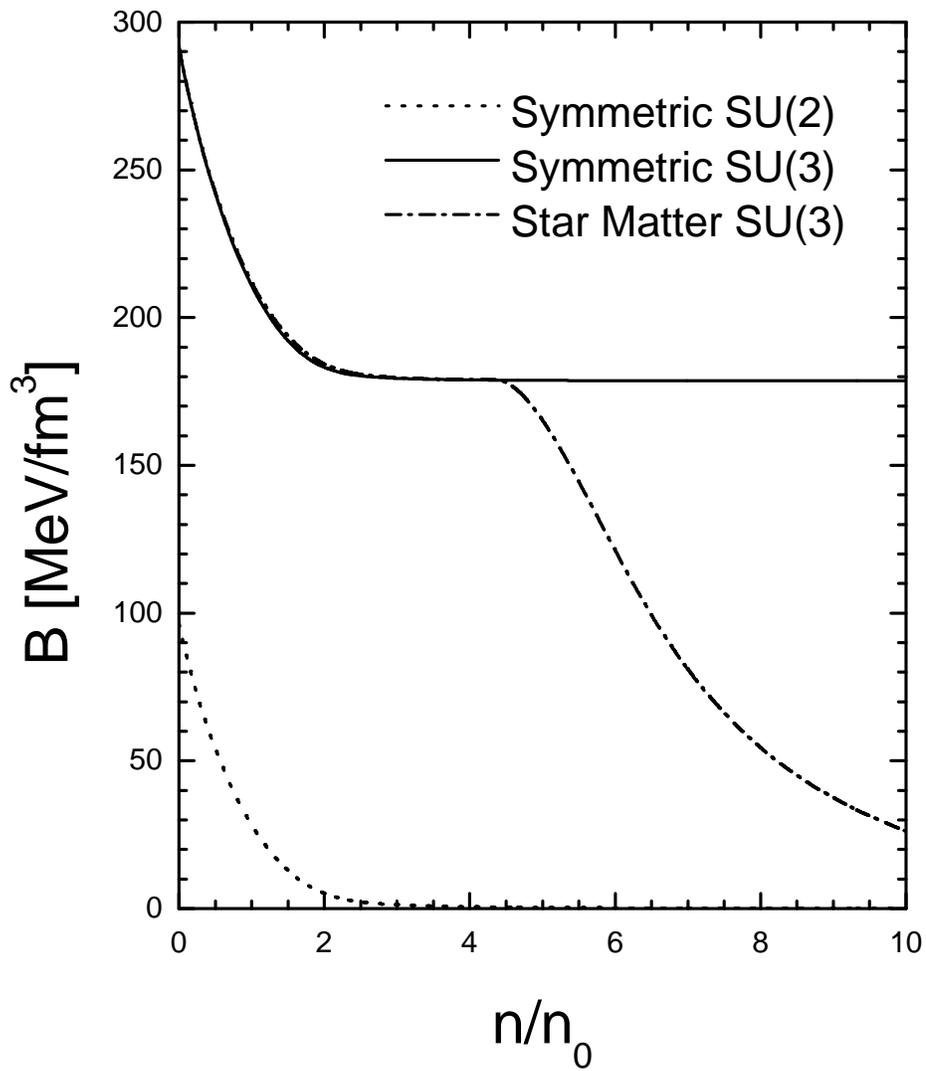} \caption{
 The bag parameter in terms of the normalized baryon number
density.}
\end{figure}

\newpage

\begin{figure}
\centering
\includegraphics[height=0.85\textheight]{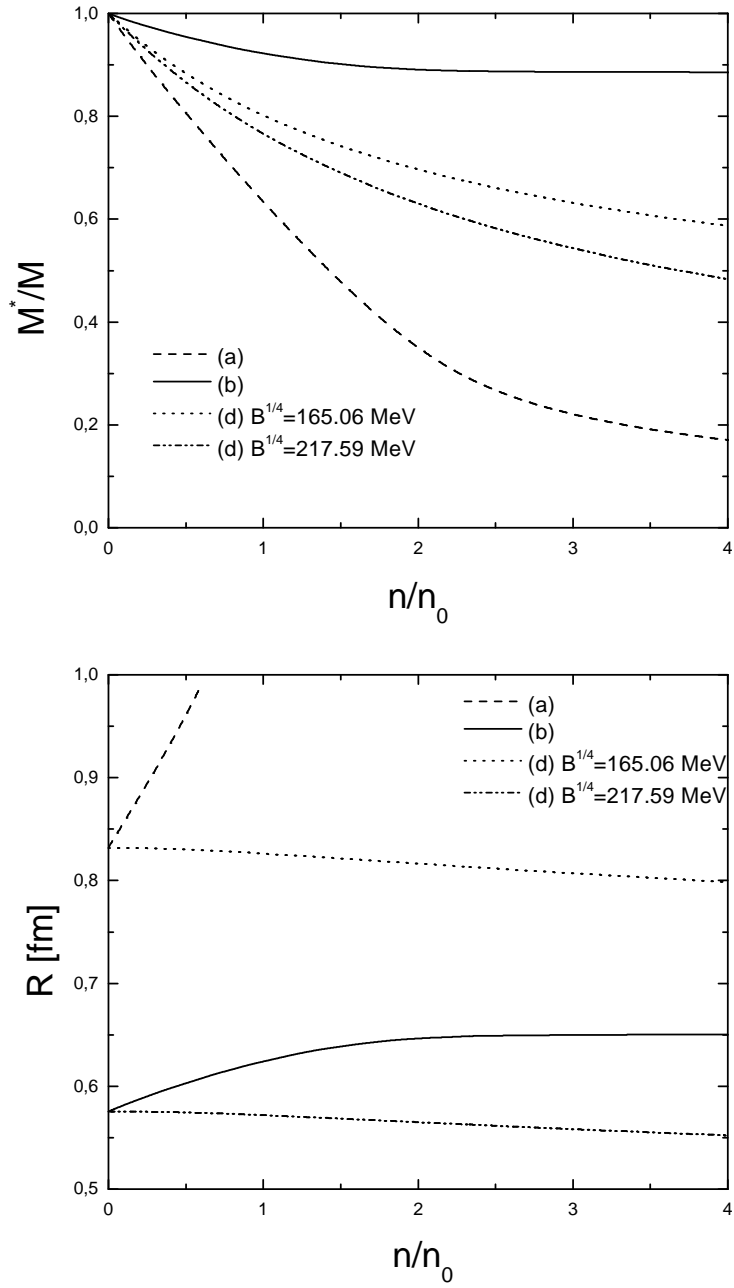}
\caption{In-medium nucleon properties as functions of the
normalized baryon number density.  In the upper figure the
effective neutron mass relative to its value at zero baryon number
density, in the lower figure the equilibrium bag radius. The line
convention is explained in each figure, the labels a,b, and d have
the same meaning as in the text. In the last case we have taken
two different constant values for $B$ obtained at zero baryon
density in the treatments (a) and (b).}
\end{figure}

\newpage

\begin{figure}
\centering
\includegraphics[height=0.85\textheight]{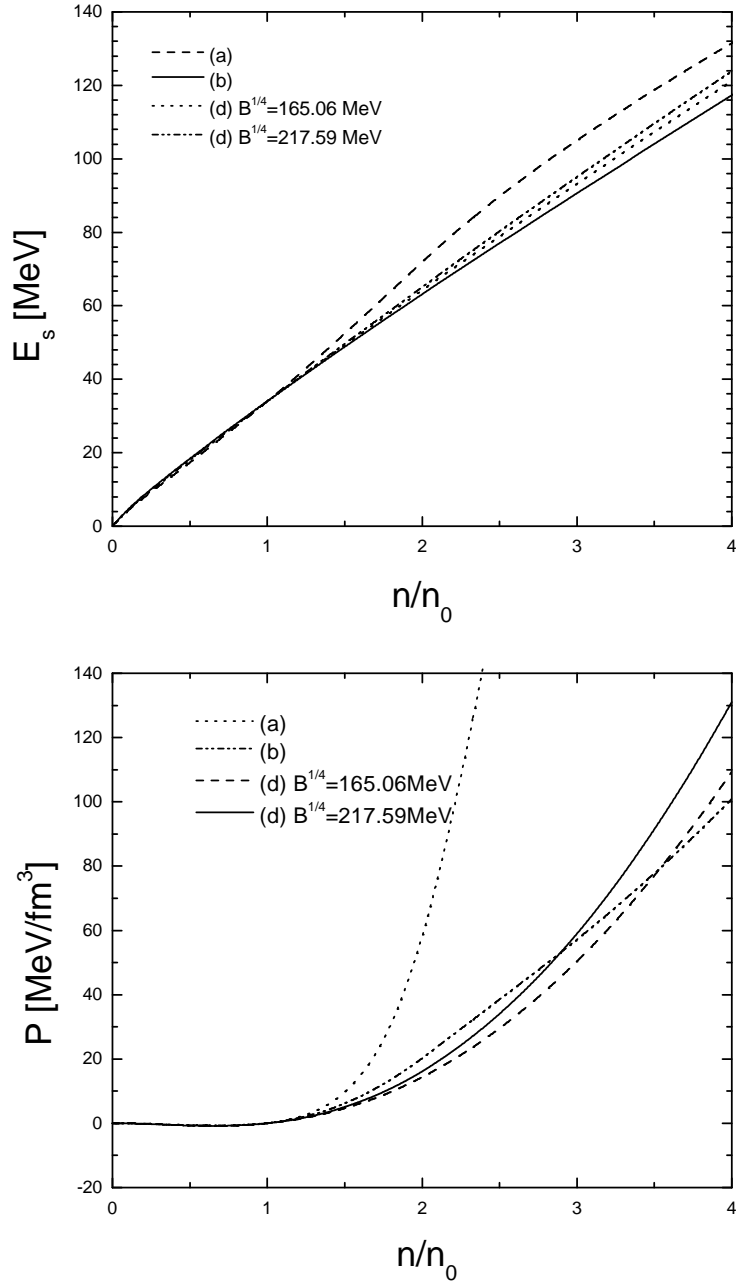}
\caption{Thermodynamical properties of symmetric nuclear matter as
functions of the normalized baryon number density.  In the upper
figure the symmetry energy and the pressure in the lower panel.
For the line convention see the explanation in Fig.2.}
\end{figure}

\end{document}